# TRANSPORT FROM THE RECYCLER RING TO THE ANTIPROTON SOURCE BEAMLINES*

M. Xiao[#], Fermilab, Batavia, IL 60540, USA

*Abstract*

In the post-NOvA era, the protons are directly transported from the Booster ring to the Recycler ring rather than the Main Injector. For Mu2e and g-2 project, the Debuncher ring will be modified into a Delivery ring to deliver the protons to both Mu2e and g-2 experiments. Therefore, it requires the transport of protons from the Recycler Ring to the Delivery ring. A new transfer line from the Recycler ring to the P1 beamline will be constructed to transport proton beam from the Recycler Ring to existing Antiproton Source beamlines. This new beamline provides a way to deliver 8 GeV kinetic energy protons from the Booster to the Delivery ring, via the Recycler, using existing beam transport lines, and without the need for new civil construction. This paper presents the Conceptual Design of this new beamline.

## INTRODUCTION

During antiproton stacking for Tevatron Collider operations, 120 GeV protons are transferred from the Main Injector to the P1 beamline. The beam is then directed to the antiproton production target by way of the P2 and AP1 beamlines. The antiproton production target can be bypassed for 8 GeV kinetic energy protons allowing beam transport directly to the Debuncher ring via the AP3 line. Thus, it is possible to transport protons from the Main Injector to the Debuncher ring using existing beamlines. These beamlines are labelled and shown schematically in Fig. 1[1].

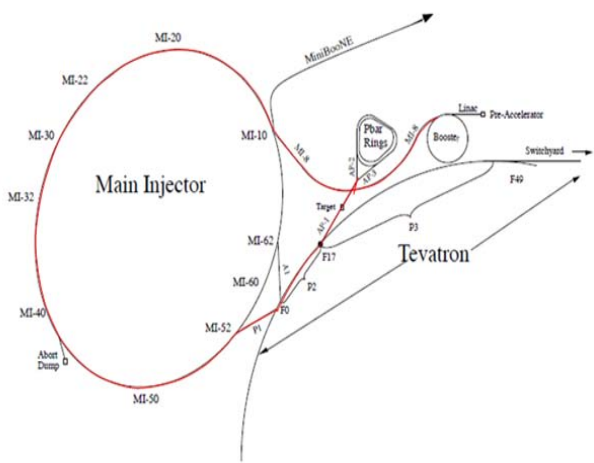

Figure 1: The Fermilab accelerator complex. The path of protons from the Booster synchrotron to the Antiproton Source Rings is shown in red.

In the post-Nova era, the protons are directly transported from the Booster ring to the Recycler ring rather than the Main Injector [2]. For Mu2e and g-2 project, the Debuncher ring will be modified into a Delivery ring to deliver the protons to both Mu2e and g-2 experiments. Therefore, it requires the transport of protons from the Recycler Ring to the Delivery ring. A new transfer line from the Recycler ring to the P1 beamline will be constructed to transport proton beam from the Recycler Ring to existing Antiproton Source beamlines. This new beamline provides a way to deliver 8 GeV kinetic energy protons from the Booster to the Delivery ring, via the Recycler, using existing beam transport lines, and without the need for new civil construction. This paper presents the Conceptual Design of this new beamline.

The P1 line is lower in elevation than the Recycler ring. Thus, a horizontal kick (traditional kicker) and a vertical bend (Lambertson magnet) will be used to extract the beam from the Recycler ring. Due to space limitations, only two bending centers are used. A vertical bending magnet is used to bend the beam into the center of the P1 line. An integer multiple of 360° in betatron phase advance between the two bending centers is required to cancel the vertical dispersion after bending down and up.

## BEAMLINE DESIGN

Fig. 2 shows the layout of the proposed transfer line from the Recycler Ring to the P1 beamline. This design extracts beam from the Recycler downstream of Recycler quadrupole Q522B and transports it to a location between quadrupoles Q702 and Q703 in the P1 beamline. The total length of the transfer line from the Recycler to the P1 line is about 43 meters.

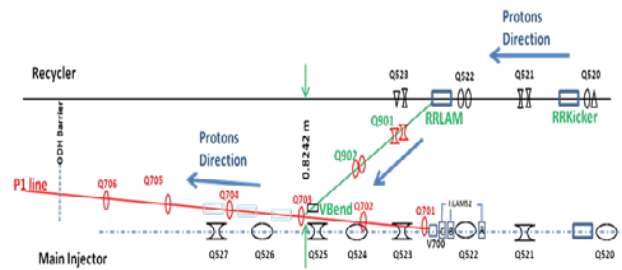

Figure 2: Schematic layout of the transfer line from the Recycler Ring to the P1 line

Table 1 lists the results for parameters of each element used in the transfer line by matching the new beamline to the P1 line at Q703. The strengths of the quadrupoles Q901A&Q901B, Q902A&Q902B were obtained using code MAD by matching the Twiss functions from the extraction point in RR and injection point in Antiproton source beamline P1. The angles of the both lambertson and the vertical bending magnet were obtained by matching the site coordinates from the RR to P1 line using the code TRANSPORT. The beam gets steered onto the nominal P1 trajectory by slightly rolling the vertical bending magnet. Fig. 3 gives the plot of the Twiss functions of the whole transfer line.

___________________________________________

*Work supported by U.S. Department of Energy under contract No. DE-AC02-76CH03000.
# meiqin@fnal.gov

Table 1: Matching results

| Element | parameter | Results (unit) |
|---|---|---|
| Q901A, Q901B | $k_1$ | -0.0885 (1/m$^2$) |
| Q902A, Q902B | $k_1$ | 0.07210 (1/m$^2$) |
| RRLAM VBEND | Bending angle θ | 20.994 mrad |
| RLROLL | Roll angle of RR lambertson | 0.0472 rad |
| VROLL | Roll angle of the vertical bending magnet | -0.0692 rad |

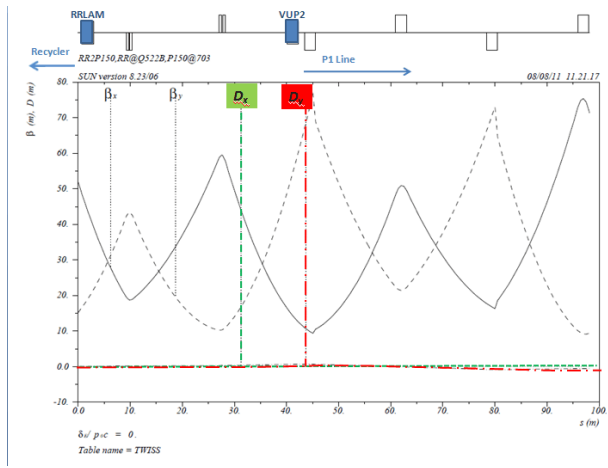

Figure 3: The Beta functions and dispersions of the new transfer line.

## THE RECYCLER EXTRACTION KICKER FOR MU2E

The Recycler extraction channel is designed such that a permanent horizontal 3-bump in the Recycler displaces the circulating beam outward 25 mm at the upstream end of the Recycler lambertson (RRLAM). The extraction kicker produces an inward horizontal displacement of 25 mm at the same location for the extracted beam. Thus, the separation of the Recycler circulating beam and the extracted beam at the front face of the lambertson is 50 mm. Fig. 4 shows the trajectories of the circulating and the extracted beam.

The three Recycler correctors used for the 3-bump of the circulating beam are H524, H522 and H524. The strengths of these correctors in radians were obtained to be:

$$\begin{cases} hk524 = 7.514E - 04 \\ hk522 = -3.865E - 04 \\ hk520 = 6.952E - 04 \end{cases}$$

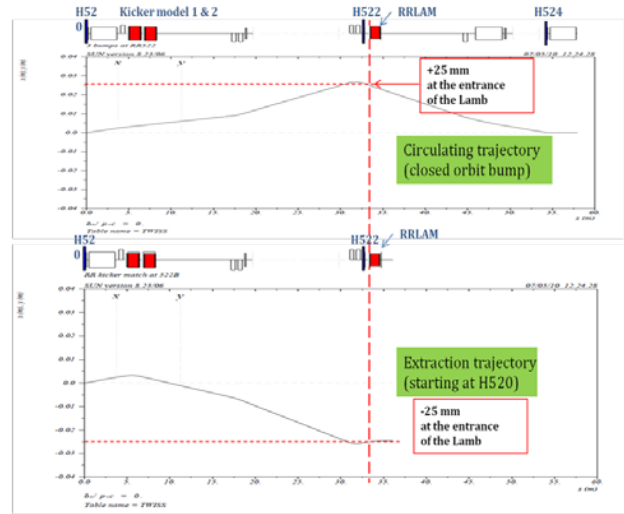

Figure 4: Trajectories of the circulating and extracted beam upstream of the Recycler extraction Lambertson.

The Recycler extraction kicker for Mu2e consists of two modules of a NOvA style kicker magnet. The extraction kicker for Mu2e will be the same design as the gap clearing kicker used in Recycler for Nova project. Seven kicker modules will be needed to extract Mu2e designated beam from the Recycler Ring. One of these modules is used to compensate for the decaying "tail" of the kicker waveform, so that circulating beam is not disturbed. The total kicker angle required is 1.585 mrad. Each of the kicker modules will contribute 0.226 mrad, which is approximately 92% of the design maximum strength. All of the power supplies for the new magnets for this beam line will be located in the MI-52 service building. This service building will be expanded to accommodate the new power supplies.

## ACCEPTTANCE AT THE LAMBERTSON AND VERTICAL BENDING DIPOLE

We have checked the acceptance of the beamline. The tightest spots would be at 2 bending magnets- the lembertson and the vertical bending dipole. Fig. 5 shows the aperture at the Lambertson magnet for the circulating beam and the extracted beam. The beam size at the center of the Lambertson is $\sigma_x$ = 3.518 mm and $\sigma_y$ = 2.627 mm given the emittance of the beam 18 πmm.mrad.

Fig. 6 shows the aperture at the vertical bending magnet (VUP). The beam size for 18 πmm.mrad beam emittance at this location is 1.989 mm × 4.377 mm. The physical aperture of a modified 5-ft (1.524 m) B1 dipole is 32.258 mm × 121.158 mm. The bending strength required is 20.996 *mrad*. The calculated *sagitta* is 16 mm. The acceptance is then: 32.258 / 1.989 = 16.2 $\sigma_x$ (horizontal) and (121.158-16.0) / 4.377=24.0 $\sigma_y$ (vertical).

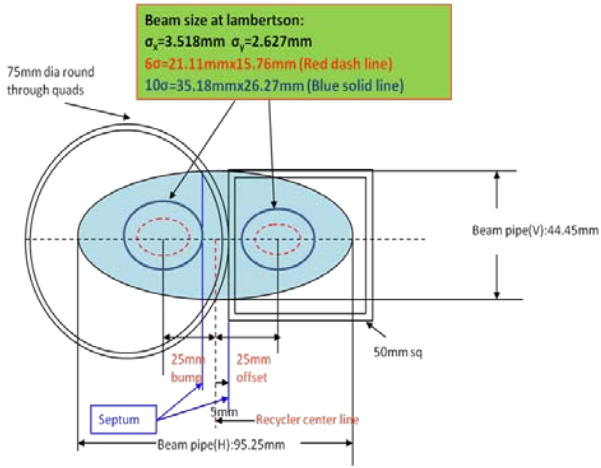

Figure 5: Acceptance at the Lambertson magnet. The emittance of the beam is given 18πmm.mrad. The Recycler's elliptical beam pipe is shown in light blue. At the left hand side of the Recycler center line, the 6σ contour of the circulating beam is shown by a dashed red line. The solid dark blue line shows the 10σ contour of the circulating beam. At right hand side, are shown the 6σ (dashed red line) and 10σ (solid dark blue line) contours of the extracted beam.

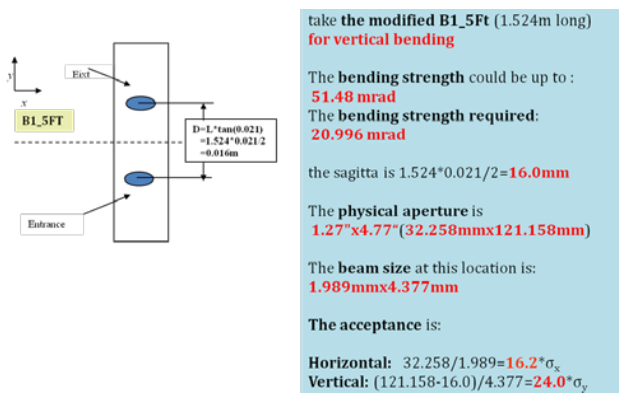

Figure 6: Acceptance of the vertical bending magnet.

## CONCLUSION

The concept design of the transfer line from the RR to Antiproton source beamline was completed and the specification of all beamline elements are summarized and given in Table 2 for cost evaluations.

## ACKNOWLEDGMENT

Many thanks to D.E Johnson for useful discussions. Special thanks to Steven Werkema for his help in report.

Table 2 Specifications of the beamline elements

| Element | Requirement |
|---|---|
| Kicker | **Angle:** 1.585 mrad, (0.470 kG-m integrated field). Need bending the beam outside of the RR, B-field is up<br>**Orientation:** Horizontal<br>**Location:** Center of kicker is 5.02 m "Down Stream" MRK 520<br>**Physical and field aperture:** 33 mm V × 81 mm H elliptical shape<br>**Field Flattop time:** 1. 534 μs (81 52.809 MHz buckets)<br>**Field Rise Time:** 57 nsec maximum<br>**Field Fall time:** 57 nsec maximum<br>**Flattop ripple and tilt**: ±4%<br>**Flattop repeatability** : ±3%<br>**Max kick to circulating beam**: ±3% of nominal<br>**Vacuum**< $10^{-8}$ Torr<br>**Power Supply & Cooling Location**: MI-52<br>**Number of injection per cycle:** 2<br>**Time between transfers** :0.067 seconds<br>**Time between machine loads** 1.33 seconds |
| Lambertson | **Bending angle**: 20.996 mrad, (6.225 kG-m integrated field),<br>(8 GeV injection Lambertson, L=2.286m, aperture: 120mm(H), 44mm(V)<br>**Roll angle** : 2.7°<br>**Power supply**: DC |
| V-bending dipole | **Bending angle**: 20.996 mrad, (6.225 kG-m integrated field) ADCW (modified B1 style to open aperture)<br>**Roll angle**: 3.97°<br>Power Supply: DC |
| Quads | 4 permanent RR type, ±10% standard strength.<br>4 trim quads for ±20% adjustable ability, power supply: DC |
| BPMs | 2 Horizontal and 2 vertical<br>8 GeV style BPM 4" aperture |
| Correctors | 1 Horizontal and 1 vertical, power supply: DC |
| Loss monitors | 2 |
| Multiwires | 2 |